# Carbon Monitor-Power: near-real-time monitoring of global power generation on hourly to daily scales

Biqing Zhu[1,2], Xuanren Song[1], Zhu Deng[1,3], Wenli Zhao[4], Da Huo[1,5], Taochun Sun[1], Piyu Ke[1], Duo Cui[1], Chenxi Lu[1], Haiwang Zhong[6], Chaopeng Hong[7], Jian Qiu[3], Steven J. Davis[8], Pierre Gentine[6], Philippe Ciais[2,*], Zhu Liu[1,*]

[1] Department of Earth System Science, Tsinghua University, Beijing, 100084, China
[2] Laboratoire des Sciences du Climate et de l'Environnement LSCE, Orme de Merisiers 91191, Gif-sur-Yvette, France
[3] Product and Solution & Website Business Unit, Alibaba Cloud, Hangzhou, Zhejiang, 311121, China
[4] Department of Earth and Environmental Engineering, Columbia University, New York, NY,USA
[5] Department of Civil & Mineral Engineering, University of Toronto, Toronto, ON, M5S 1A1, Canada
[6] Department of Electrical Engineering, the State Key Lab of Control and Simulation of Power Systems and Generation Equipment, Institute for National Governance and Global Governance, Tsinghua University, Beijing, China
[7] Institute of Environment and Ecology, Shenzhen International Graduate School, Tsinghua University, Shenzhen, China.
[8] Department of Earth System Science, University of California, Irvine, 3232 Croul Hall, Irvine, CA, 92697-3100, USA
* Corresponding authors: Zhu Liu (zhuliu@tsinghua.edu.cn), Philippe Ciais (philippe.ciais@cea.fr)

## Abstract

We constructed a frequently updated, near-real-time global power generation dataset: Carbon Monitor-Power since January, 2016 at national levels with near-global coverage and hourly-to-daily time resolution. The data presented here are collected from 37 countries across all continents for eight source groups, including three types of fossil sources (coal, gas, and oil), nuclear energy and four groups of renewable energy sources (solar energy, wind energy, hydro energy and other renewables including biomass, geothermal, etc.). The global near-real-time power dataset shows the dynamics of the global power system, including its hourly, daily, weekly and seasonal patterns as influenced by daily periodical activities, weekends, seasonal cycles, regular and irregular events (i.e., holidays) and extreme events (i.e., the COVID-19 pandemic). The Carbon Monitor-Power dataset reveals that the COVID-19 pandemic caused strong disruptions in some countries (i.e., China and India), leading to a temporary or long-lasting shift to low carbon intensity, while it had only little impact in some other countries (i.e., Australia). This dataset offers a large range of opportunities for power-related scientific research and policy-making.

| | |
|---|---|
| Measurement(s) | Power generation by different energy sources |
| Technology Type(s) | computational modeling technique |
| Factor Type(s) | geographic location • energy sources • temporal interval |
| Sample Characteristic - Environment | power system |
| Sample Characteristic - Location | global |

## Background & Summary

Power is a fundamental element of human society. Access to affordable, reliable, and sustainable energy, including access to reliable electricity and power produced by renewable sources, are listed as important aspects of the United Nations Sustainable Development Goals[1]. Tracking dynamics and status of power production and consumption is of great importance as it reflects the manufacturing, social activities as well as human impacts on the environment. Current power statistics are based on inventories of power production, consumption, trade, etc[2]. This work usually has a time lag of at least one year[3–7]. Timely and effective management of the power sector, including monitoring shifts from fossil to low carbon sources, is valuable for effectively mitigating global climate change policy-making[8,9]. Thus low-latency data on global and national power production with the high-temporal resolution is urgently needed[10].

As a result, high-temporal resolution power data is increasingly important and has received a focus from increasing from governments, companies, and academic institutes[11–13]. Daily and hourly power data are critical to developing power system models[14], or to understanding the patterns of human behaviors[15,16]. With the increasing awareness of the importance of such datasets, there has been an increase in open access at regional levels. For example, the EU has created an open platform ENTSO-E for electricity generation, load, and transmission data for Europe[12]. United States' Energy Information Administration (EIA) also provides free access to data for its electricity generation and consumption[13]. China's electricity generation and consumption data are available through its national grid or China's National Bureau of Statistics[17]. However, the temporal coverage often varies between datasets. In addition, energy sources are reported or aggregated differently[11,17–20]. These inconsistencies have made it challenging to compare and evaluate progress in decarbonizing power systems across countries and regions.

International Agency (IEA) and BP provide well-integrated and unified data for power generated from different sources and cover a wide range of spatial regions[2,20]. International Renewable Agency (IRENA) also provides reports on global renewable energy installed capacity and generation[19]. However, those datasets have a time lag of at least several months and have at best a monthly time step. Monthly datasets may not provide sufficient information on power systems' rapid changes, due to 1) changes in human behavior as the COVID-19 pandemic or the effects of weekends, and holidays[21–24], 2) the impact of climate variabilities such as winter storms, summer heatwaves, and other climate variabilities causing shifts of demand, and intermittency of renewable power supply[25], and 3) economic shocks such as abrupt variations of fuel prices or shortfalls of supply since the war between Ukraine and Russia[26].

Here we constructed the first global daily and hourly power generation dataset (Carbon Monitor-Power) for the period going from 2016-01 to 2022-07. This dataset can be updated in near-real-time with a latency of between 1 day to a maximum of 1 month, depending on the country/region. The dataset includes daily and hourly power generation data from fossil fuels (coal, natural gas, and oil) and nuclear, hydro, wind, solar, geothermal, biomass, and other renewables for 37 countries, which covers around 70% of the global power production and 68% of global power-related $CO_2$ emissions. Carbon Monitor-Power provides a data basis to the Carbon Monitor dataset, to estimate the near-real-time daily $CO_2$ emissions from power generation[21–23]. Carbon Monitor-Power represents a new resource for exploring high-time frequency patterns of the global power system and monitoring monthly to annual changes relevant to emissions reduction pledges (**Fig 1**).

# Methods

We mostly collected data from national grid operators which provide open-access power generation and consumption data at high temporal frequency. In constructing a harmonized database with global coverage, special attention was given to filling the data gaps. Although it is possible to directly acquire high-time-frequency power generation data for the EU for example, such data does not exist for some other countries like China. China's national grid provides detailed information on installed capacity and utilization hours for major power sources, but every month.

The framework used to generate the Carbon Monitor-Power dataset is shown in **Fig 2**. We acquired raw data from the national grids of the 37 countries/regions listed in Table 1. Raw data are acquired at the highest possible time resolution (5-minute intervals, hourly, daily, or monthly, depending on the source availability). We then developed national-specific methods for data processing and simulation (details see country-specific method below).

## Power Generation Data Acquisition

We firstly collected raw data from 12 regions (or 37 countries, including Australia, Brazil, China, 26 countries in EU27, UK, India, Japan, Russia, South Africa, the United States, Mexico, and Chile) with various energy sources. The raw data are collected from publicly available sources at the national or subnational levels. Data sources used in this study are summarized in **Table 1**.

More than two million records of raw data have been collected from these 12 data sources, with nearly two thousand records newly generated and collected per day. Considerable data cleaning preprocess was performed as part of the data processing, due to frequent extreme values and missing values detected from near-real-time data.

To filter out extreme values, we first examine the quality of these high-temporal power generation datasets using an Interquartile Range (IQR) threshold method[27] and detect the 'outliers'. The IQR range is defined as the range between the $75^{th}$ percentile and the $25^{th}$ percentile. The upper limit is calculated as adding .5*IQR to the $75^{th}$ percentile. The lower limit is calculated as subtracting 1.5*IQR from the $25^{th}$ percentile. Values fall beyond the upper and lower limits are labeled as potential 'outliers. Afterward, manual processing was applied to evaluate whether each extreme value should be removed or to be kept. As a general rule, we keep extremes in the data set when there was evidence of abrupt social changes (COVID-19 confinements) and/or natural disasters (e.g., storms), which are known to have a strong and sudden impact on the power system. Then the linear interpolation function from the Python Pandas package was used to fill missing values.

After such detections, no outliers or missing values have been found in the raw data from most countries, even some of which may have pre-processed their raw data before releasing them publicly (such as the United States EIA[13]). In the end, pre-processing of removing outliers and filling missing values was only conducted on the raw data from China and EU27. Data records which are identified as outliers or missing values are labeled as F (Filtered). Others are labeled as N (Normal)

## Country-specific data process

After the data preprocessing, for each country/region, we aggregate and/or dis-aggregate the power generation to daily (or hourly if possible) according to data availability, and to eight categories of power generation sources: coal, natural gas, petroleum, nuclear, hydro-power, wind, solar and other renewable (including biomass, geothermal, and power generated from residual industrial heat and from other non-specified sources). In addition to those national data, three additional data sets are used to disaggregate the total generation data into specific generation types when needed: Monthly Electricity Statistics by IEA[20], Statistical Review of World Energy by BP[2], and the Renewable Energy Statistics by IRENA[19].

However, the original data often do not have the same format and do not cover the same energy sources across countries. To establish a harmonized dataset with all sources of power, we collected information from other databases with a lower time frequency and disaggregated them to daily steps (details see country-specific method below).

## Australia

The original data is acquired at hourly resolution for the following categories: *Wind*, *Hydro*, *Solar (Rooftop)*, *Solar (utility)*, *Gas (Waste Coal Mine)*, *Gas (Reciprocating)*, *Gas (OCGT)*, *Gas (CCGT)*, *Gas (Steam)*, *Distillate (Energy source: Diesel)*, *Bioenergy (Biomass)*, *Bioenergy (Biogas)*, *Coal (Black)*, *Coal (Brown)*, while, *CCGT* refers to power generated by combined cycle gas turbine, and *OCGT* refers to power generated by open gas cycle turbine. We aggregate all power generated from all types of power used in this study as following:

$$P_{coal,h} = P_{Coal.Black,h} + P_{Coal.Brown,h} \quad (1)$$

$$P_{gas,h} = P_{Gas.steam,h} + P_{Gas.CCGT,h} + P_{Gas.OCGT,h} + P_{Gas.Reciprocating,h} + P_{Gas.Waste\ Coal\ Mine,h} \quad (2)$$

$$P_{solar,h} = P_{Solar.Utility,h} + P_{Solar.Rooftop,h} \quad (3)$$

$$P_{other\ renewable,h} = P_{Other.Bioenergy\ biomass,h} + P_{Other.Bioenergy\ biogas,h} \quad (4)$$

## Brazil

The raw power generation data from Brazil is acquired from the Operator of the National Electricity System (http://www.ons.org.br/Paginas/). The data acquisition and download are performed at a daily base, with up to a week of latency (occasional delay caused by site maintenance). The original data is acquired at hourly resolution for the following categories: *Wind*, *Hydro*, *Nuclear*, *Solar*, *Thermal* (including *Coal*, *Coal Mineral*, *Gas*, *Natural Gas*, *Combustive Oil*, *Diesel*, *Petrol (Gasoline)*, *Biomass*, *Industrial Residuals*). We aggregate power generated from all types of coals to coal power, and similarly, power generated from all gas types to gas power. The further aggregation to the power generated by energy source ($s$) at each hour ($P_{s,h}$) used in this study is as following:

$$P_{oil,h} = P_{Combustive\ Oil,h} + P_{Diesel,h} + P_{Petrol,h} \quad (5)$$

$$P_{other\ renewable,h} = P_{Biomass,h} + P_{Industrial\ Residuals,h} \quad (6)$$

The power generation data is firstly produced at hourly time resolution, then further aggregated to daily resolution.

## China

There are two types of core datasets for China power generation: Power Generation by Energy Type ($P$) and Coal Consumption data ($CC$). The Power Generation by Energy Type data is acquired from China's Electricity Council (CEC, http://cec.org.cn), which provides information on power generation, consumption and usage on China's national grid at monthly, seasonal and yearly time steps. The information on power generation is given primarily as installed capacity per energy source ($IC$), and cumulative utilization hour ($CUH$) per type of source ($s$). The power generation for month m from energy source s ($P_{s,m}$) is calculated as:

$$P_{s,m} = IC_{m,s} \times (CUH_{m,s} - CUH_{m-1,s}) \quad (7)$$

This allows the direct calculation of power generation from energy sources including thermal, coal, natural gas, nuclear, hydro, wind, solar, biomass and geothermal (the last two with compromised time frequency and latency). Power generation by oil and other non-fossil sources, such as waste, recovery energy from industrial processes is not provided separately by CEC. But they are accounted for in the total thermal power. Therefore, we separated the thermal power from coal and gas production using factors derived from Monthly Electricity Statistics by IEA ($P_{IEA}$) from the corresponding month of the latest year available:

$$F_{oil,m} = \frac{P_{IEA,Oil,m}}{P_{IEA,Combustible\ Renewables,m} + P_{IEA,Other\ Combustibles,m} + P_{IEA,Oil,m}} \quad (8)$$

$$F_{other\ thermal,m} = \frac{P_{IEA,Combustible\ Renewables,m} + P_{IEA,Other\ Combustibles,m}}{P_{IEA,Combustible\ Renewables,m} + P_{IEA,Other\ Combustibles,m} + P_{IEA,Oil,m}} \quad (9)$$

The power generated by oil and by other renewable energy are then simulated as:

$$P_{oil,m} = (P_{thermal,m} - P_{coal,m} - P_{gas,m}) \times F_{oil,m} \quad (10)$$

$$P_{other\ renewable,m} = (P_{thermal,m} - P_{coal,m} - P_{gas,m}) \times F_{other\ thermal,m} \quad (11)$$

The monthly power generation data are then disaggregated to daily values using daily coal consumption ( $CC$ ) data by power plants from eight coastal provinces in China (https://www.cctd.com.cn/) with the following equation:

$$P_{s,d} = P_{s,m} \times \frac{CC_d}{CC_m} \quad (12)$$

Where $P_{s,d}$ being the power generation by energy source $s$ on day $d$, $P_{s,m}$ being the power generation of the month $m$. $CC_d$ is the coal consumption on day $d$, and $CC_m$ is the monthly coal consumption of month $m$. Day $d$ is within month m. The correlation between China's power generation and coal consumption was established at monthly time steps, and the detailed method can be found in our previous work[21,22].

## EU27&UK

For the year between 2016 and 2018, the power generation data for EU28 are acquired from ENTSO-E (https://www.entsoe.eu). From 2018 onwards, data for EU27 was continuously acquired from ENTSO-E, while data for the UK was acquired from BMRS (https://www.bmreports.com/bmrs/) and from Sheffield Solar (for solar power only, https://www.solar.sheffield.ac.uk/). The data acquisition is carried out on a daily basis. The data aggregation method is database specific instead of country specific.

ENTSO-E provides power generation capacity data at time resolution between every 15 minutes to every 60 minutes. Therefore, hourly power generation was firstly computed as

$$P_{c,s,h} = Capacity_{c,s,h} \times 1hr \quad (13)$$

The power generation ($P$) from country $c$ from source $s$ during the hour $h$ is computed as the average power generation capacity during the corresponding time, multiplied by the length of duration (1 hour). Then the power generation data for each country ($c$) was further aggregated to the eight energy sources used in this study with the following equations at desired time resolution (excluding energy sources where aggregation was not required):

$$P_{c,coal,h} = P_{c,B\ rown\ coal\ \&Lignite,h} + P_{c,Coal\ derived\ gas,h} + P_{c,Hard\ coal,h} \quad (14)$$

$$P_{c,oil,h} = P_{c,Oil,h} + P_{c,Shale\ Oil,h} \quad (15)$$

$$P_{c,hydro,h} = P_{c,Pumped\ Storage,h} + P_{c,Run\ of\ river\ and\ poundage,h} + P_{c,Water\ reservoir,h} \quad (16)$$

$$P_{c,wind,h} = P_{c,Wind\ Offshore,h} + P_{c,Wind\ Onshore,h} \quad (17)$$

$$P_{c,other,h} = P_{c,BIomass,h} + P_{c,Geothermal,h} + P_{c,Other\ reneable,h} + P_{c,Waste,h} \quad (18)$$

BMRS provides power generation capacity data over the UK for 48 equally distributed time periods per day. The conversion from power generation capacity to hourly power generation follows the same methods as equation 13. Two energy sources were aggregated from the database to match the sources defined in this study, with the following equations:

$$P_{gas,h} = P_{ccgt,h} + P_{ocgt,h} \quad (19)$$

$$P_{hydro,h} = P_{Hydro(Pumped\ storage),h} + P_{Hydro(Non-pumped\ storage),h} \quad (20)$$

$$P_{other,h} = P_{Biomass,h} + P_{Other,h} \quad (21)$$

Among which, $ccgt$ refers to power generated by combined cycle gas turbine, and $ocgt$ refers to power generated by an open gas cycle turbine. The sources aggregated hourly power generation data was then further aggregated to daily, monthly and yearly resolution. In the end, we aggregate all countries to EU27 and UK to an aggregated dataset for EU27&UK.

## India

The power generation data from India was initially acquired from the Power System Operation Corporation Limited (POSOCO, https://posoco.in/) on a daily basis, with one to two days of latency. The original data is provided for aggregated sources as compared to our required eight sources (**Fig 2**): Gas and Oil produced power are aggregated and called *Gas_Naptha_Diesel*; *RES* aggregates power produced by wind, solar, biomass, and other energy sources. To disaggregate these energy sources, we developed factors ($F$) based on the reference monthly power generation dataset ($P_{IEA}$)[20] of the corresponding month ($m$) from the last available year (the current year or the previous year). for oil and gas:

$$F_{s,m} = \frac{P_{IEA,s,m}}{P_{IEA,Oil,m} + P_{IEA,Natural\ Gas,m}} \quad (22)$$

For solar and wind:

$$F_{s,m} = \frac{P_{IEA,s,m}}{\sum_m P_{IEA,Solar}, P_{IEA,Wind}, P_{IEA,Combustible\ Renewables}, P_{IEA,Geothermal}, P_{IEA,Other\ Renewables})} \quad (23)$$

For other renewable:

$$F_{s,m} = \frac{\sum_m (P_{IEA,Combustible\ Renewables}, P_{IEA,Geothermal}, P_{IEA,Other\ Renewables})}{\sum_m (P_{IEA,Solar}, P_{IEA,Wind}, P_{IEA,Combustible\ Renewables}, P_{IEA,Geothermal}, P_{IEA,Other\ Renewables})} \quad (24)$$

With these factors, we disaggregate the daily power production from its original aggregated sources as follows:

$$P_{gas,d} = P_{Gas_Naptha_Diesel,d} \times F_{gas,m} \quad (25)$$

$$P_{oil,d} = P_{Gas_Naptha_Diesel,d} \times F_{oil,m} \quad (26)$$

$$P_{solar,d} = P_{RES,d} \times F_{solar,m} \quad (27)$$

$$P_{wind,d} = P_{RES,d} \times F_{wind,m} \quad (28)$$

$$P_{other\ renewables,d} = P_{RES,d} \times F_{other\ renewables,m} \quad (29)$$

With day $d$ being in the month $m$.

## Japan

We acquire Japan's hourly power generation data from the Organization for Cross-regional Coordination of Transmission Operators (OCCTO, https://www.occto.or.jp/en/), with a latency of one to two months. The data provides fossil power in one aggregated sector. Therefore, we disaggregate fossil power with factors derived from reference monthly power generation data[28] for the coal, gas, and oil sectors:

$$P_{s,h} = P_{Fossil,h} \times F_{s,m} \tag{30}$$

$$F_{s,m} = \frac{P_{s,m}}{\sum_m (P_{IEA,Coal},\ P_{IEA,Natural\ Gas},\ P_{IEA,Oil})} \tag{31}$$

Hour $h$ being on day $d$ in the month m. While for the hydro, wind, solar and other renewable categories, we apply the following aggregation:

$$P_{hydro,h} = P_{Hydroelectric,h} + P_{Pumped\ Storage\ Hydroelectricity,h} \tag{32}$$

$$P_{solar,h} = P_{Photovoltaic,h} + P_{Photovoltaic\ Regulated,h} \tag{33}$$

$$P_{wind,h} = P_{Wind,h} + P_{Wind\ Regulated,h} \tag{34}$$

$$P_{other,h} = P_{Biomass,h} + P_{Geothermal,h} \tag{35}$$

## Russia

Hourly power generation data from Russia was acquired from the United Power System of Russia (http://www.so-ups.ru/index.php). It provides power generation hourly power generation for *Thermal*, *Nuclear*, *Hydro*, and *Renewables*. We develop factors ($F_s$) based on the reference yearly power generation dataset ($P_{bp}$, from BP Statistical Review of World Energy[2]) from the last available year (being previous year or the year before, $y$). for fossil energy sources ($s$):

$$F_s = \frac{P_{bp,s,y}}{P_{bp,Coal,y} + P_{bp,Gas,y} + + P_{bp,Oil,y}} \tag{36}$$

With these factors, we disaggregate the hourly fossil power generation to coal, gas, and oil power as follows:

$$P_{s,h} = P_{\text{Thermal,h}} \times F_s \tag{37}$$

For wind and other renewables, we develop $F_s$ based on reference yearly renewable power generation dataset ($P_{IRENA}$, from International Renewable Energy Agency[19]) from the last available year (being previous year or the year before, $y$):

$$F_s = \frac{P_{IRENA,s,y}}{P_{IRENA,Wind,y} + P_{IRENA,Other,y}} \tag{38}$$

With these factors, we disaggregate the hourly power generation of the category *Other* to and other renewable power as following:

$$P_{s,h} = P_{\text{Other,h}} \times F_s \tag{39}$$

## United States

Hourly power generation data from the United States was acquired from EIA (https://www.eia.gov/beta/electricity/gridmonitor/). As the energy sources provided by EIA match the source categories used in this study, we did not apply further data aggregation steps following data acquisition and data cleaning. The data are acquired at the local time and aggregated to national totals.

## South Africa

The hourly power generation data from South Africa is acquired from Eskom (https://www.eskom.co.za/dataportal/supply-side/station-build-up-for-the-last-7-days/).
Eskom provides power generation data for the following categories: *Coal* (labeled as *Thermal* in the source data), *Natural-gas*, *Oil* (labeled as *OCGT* in the source data), *Nuclear*, *Pumped Water Generation*, *Hydro Water Generation*, *Photovoltaic generation (PV)*, *Concentrated Solar Power generation (CSP)*, *Wind*, and *Other Renewable*. We aggregate hydro power and solar power sources with the following equations:

$$P_{hydro,h} = P_{Pumped\ Water\ Generation,h} + P_{Hydro\ Water\ Generation,h} \tag{40}$$

$$P_{solar,h} = P_{South\ Africa,PV,h} + P_{CSP,h} \tag{41}$$

## Mexico

The hourly power generation data from Mexico is acquired from Gobierno De Mexico (https://www.cenace.gob.mx/Paginas/SIM/Reportes/EnergiaGeneradaTipoTec.aspx) for the following categories: *Coal*, *Gas*, *Combined cycle*, *Internal Combustion* (major fuel types including hydrocarbons such as paraffinic, olefinic, naphthenic, aromatic), *Conventional Thermal*, *Nuclear*, *Hydro*, *Wind*, *Solar*, *Biomass*, and *Geothermal*. We aggregate gas, oil power and other renewable power sources with the following equations:

$$P_{gas,h} = P_{Gas,h} + P_{Combined\ cycle,h} \tag{42}$$

$$P_{oil,h} = P_{Internal\ Combustion,h} + P_{Conventional\ Thermal,h} \tag{43}$$

$$P_{other\ renewable,h} = P_{Biomass,h} + P_{Geothermal,h} \tag{44}$$

## Chile

The hourly power generation data from Chile is acquired from Coordinador Eléctrico Nacional (https://www.coordinador.cl/operacion/graficos/operacion-real/generacion-real/). Coordinador Eléctrico Nacional provides power generation data for the following categories: *Coal*, *Petcoke*, *Biogas*, *Natural Gas*, *Diesel*, *Fuel Oil*, *Run-of-river*, *Storage*, *Wind*, *Solar*, *Geothermal*, *Biomass*, and *Cogeneration*. The coal, gas, oil, hydro and other renewables power generation are calculated with below equations:

$$P_{coal,h} = P_{Coal,h} + P_{Petcoke,h} \tag{41}$$

$$P_{oil,h} = P_{Diesel,h} + P_{Fuel\ oil,h} \tag{42}$$

$$P_{hydro,h} = P_{Run-of-river,h} + P_{Chile,storage,h} \tag{43}$$

$$P_{other\ renewable,h} = P_{Geothermal,h} + P_{Biomass,h} + P_{Cogeneration,h} + P_{Biogas,h} \tag{44}$$

## Data Records

Currently, there are 37 data records provided in this dataset, which can be downloaded at https://github.com/KowComical/CM_Power_Data or *figshare*[29]. All data are available for 1857 days (from January 1st, 2016 to 30th June, 2022 for countries except for China and South Africa. For China, the data record starts from 1st January 2018, and for South Africa from 1st April 2018):

- A record of 4,015 records are the daily total and source-specific power generation from 8 power sources (i.e., coal, natural gas, oil, hydro-power, solar-power, wind-power, biomass, geothermal and other renewable) and for 2 individual countries/regions (China (from January 2018), India).
- A record of 2,415,102 records are the hourly total and source-specific power generation from 8 power sources (i.e., coal, natural gas, oil, hydro-power, solar-power, wind-power, biomass and other renewable) and for 35 individual countries/regions (i.e., US, EU27 & UK, Russia, Japan and Brazil, Australia, Chile, Mexico, South Africa)
- Two label types are used: N (Normal) and F (Filtered). F stands for data records where the raw data are filtered out for outliers and missing values, then subsequently filled with methods stated in Power Generation Data Acquisition of the Method section. N stands for data records where no outliers and missing values are detected.

## Technical Validation

### Correlation with reference data

We compared our dataset with the reference database (IEA monthly electricity generation data[20] and BP annual electricity data[2]) over the overlapping time period of 2019 to 2022, and the results show that our data in general agrees well with the reference data (**Fig 3**). For most countries, we used the monthly electricity generation as the reference database. For countries that are not covered by IEA (Russia and South Africa), we used the BP Statistical Review of World Energy[2]. For countries compared to the IEA database, the overlapping period is 2019 to April 2022. For countries compared to BP's database, the overlapping period is 2019 to 2021. The data are displayed as monthly averaged power generation. In general, for annual total power generation, the Carbon Monitor-Power database shows good agreement with the reference dataset for all countries ($R^2 > 0.95$). This indicates that on aggregated terms, the power generation provided by Carbon Monitor-Power is in line with the reference databases. There are also strong correlations between these two databases for electricity generated by major energy sources, including coal, gas, nuclear, hydro, solar, and wind. These account for about 95.4% of total electricity generation.

There are two energy sources, however, that show lower performance, namely power generated from oil and other renewables. The lowest correlation coefficient is observed in oil-fired power ($R^2$ = 0.661). The data in **Fig 3** shows that we have a systematic overestimation for China and a systematic underestimation for Brazil for oil-generated power. For other countries, the data points are distributed with a scatter but without a bias. As oil is the smallest energy source in the power sector, it has the largest uncertainty. As for the "biomass, geothermal and other renewables" sector, the low correlation coefficient ($R^2$ = 0.682) with other databases is mainly driven by the higher estimation in China by Carbon Monitor-Power. This is mainly related to the method used: subtracting coal and gas-fired power from thermal power and then distributing the non-coal and non-gas thermal power with a disaggregation factor. We have noticed that the non-coal and non-gas thermal power collected by this database is much higher than the data provided by IEA. This led to the result that we estimate higher power produced by oil and by other renewables.

**Systematic bias with reference datasets for most countries**

The time series of Carbon Monitor-Power are compared to other datasets in **Fig 4**. The main result is that the Carbon Monitor-Power data agree well with the reference dataset in terms of the overall trend, yet with significantly shorter latency and higher temporal frequency. For most of the countries, apart from China, Carbon Monitor-Power shows lower values than the other dataset, ranging from 1% in Ember to 7% in IRENA (detailed comparison for different energy sources see **Table 2**. This is likely due to differences in input data sources. Carbon Monitor-Power uses national grids as the main data sources. Reference data sources use a combination of data sources, including national census data and self-derived estimation methods. It is very likely that most countries have off-grid power generation, which is accounted for by IRENA[19], IEA[20] and BP[30], and ember's[18] methods, but not by this study (we focus on the power grid, as stated in the method section). The exception of China is caused by the same reason. For China's power generation, we acquired our raw data from China's Electricity Council, which provides all the power distributed by the national electricity grid. The reference databases such as IEA acquired their data from China's National Bureau of Statistics, which provides power generated by Power plants above the designated size (measured by annual income). Therefore, it is likely that the power generated by small power plants and some distributed photovoltaics are not accounted for by the IEA database. But in general, the discrepancy between IEA and Carbon Monitor-Power for China is not large (except for solar power generation).

**Case study: UK's grid data**

Due to the lack of high-temporal resolution data, it is difficult to comprehensively compare all countries' energy sources with other daily or hourly power generation data. It was nevertheless possible to find one additional data source (UK_ep https://electricityproduction.uk/from/all-sources/?t=10y ) for the United Kingdom. Therefore, we compared the two databases for power generation from all eight sources at high time frequency as a case study (**Fig 5**). From the comparison, we could find that although with some discrepancies, The hourly profiles agree very well between the two data sources except for Solar power, which is not covered by the UK-ep database. This shows that the Carbon Monitor-Power database also provides reliable information at high time frequency in addition to satisfying accuracy at aggregated time steps.

## Code Availability

The generated datasets and the codes for producing the datasets are available from https://github.com/KowComical/CM_Power_Data and *figshare*[29]. The most up-to-date, continuously updated data can be visualized and uploaded from https://power.carbonmonitor.org. Codes are available upon reasonable requests.

## Acknowledgments

## Author contributions statement

B.Z, P.C. and Z.L. designed the research and wrote the manuscript. B.Z., and Z.D. designed the methods. B.Z., X.S. and Z.D conducted the data processing. T.S., P.K., D. C., and C. L. contributed to the data pre-processing. All authors contributed to data collection, discussion, and analysis.

## Competing interests

The authors declare no competing interests.

# Figures

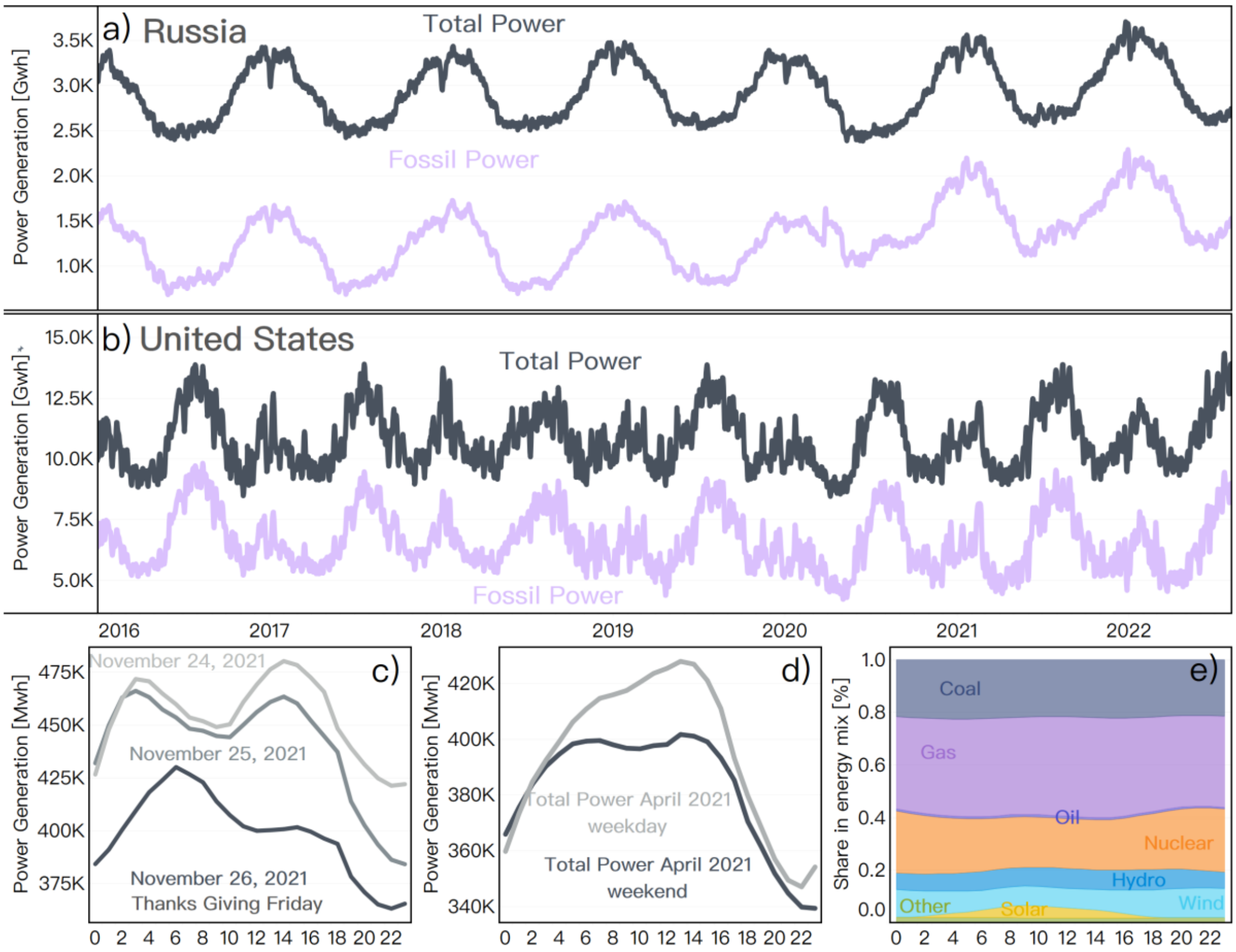

**Fig 1 | Examples of near-real-time source-specific power generation data.** a) Daily dynamics of total generation and fossil fuel generation in Russia and b) in the United States. c) Effects of holidays on diurnal profile - a strong decline of power generation in the United States during Thanksgiving Friday in 2021. d) The averaged diurnal profile of total generation of April 2021 in the United States, shows that power generation on the weekdays is higher especially during peak hours than on weekends. e) The average April diurnal profile of the energy mix of the United States' power system shows that during noon time solar power and renewables have a significantly higher share in the power system. The Carbon Monitor-Power dataset not only records the dynamics of power generation data but also provides information on the energy structure at the hourly and daily level, giving insight into the progress of decarbonization of the power system at high frequency.

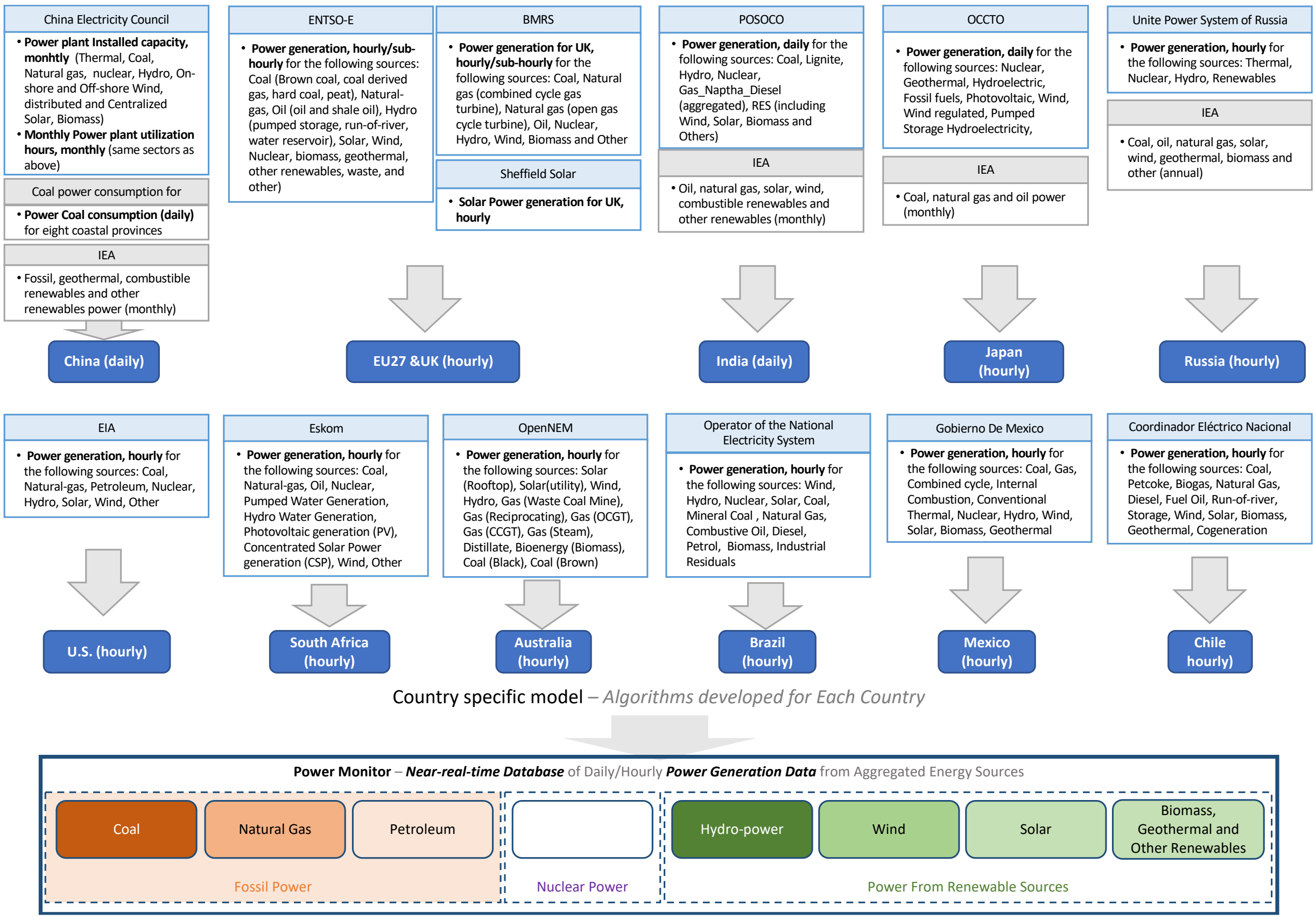

**Fig 2 | Data acquisition and processing framework**

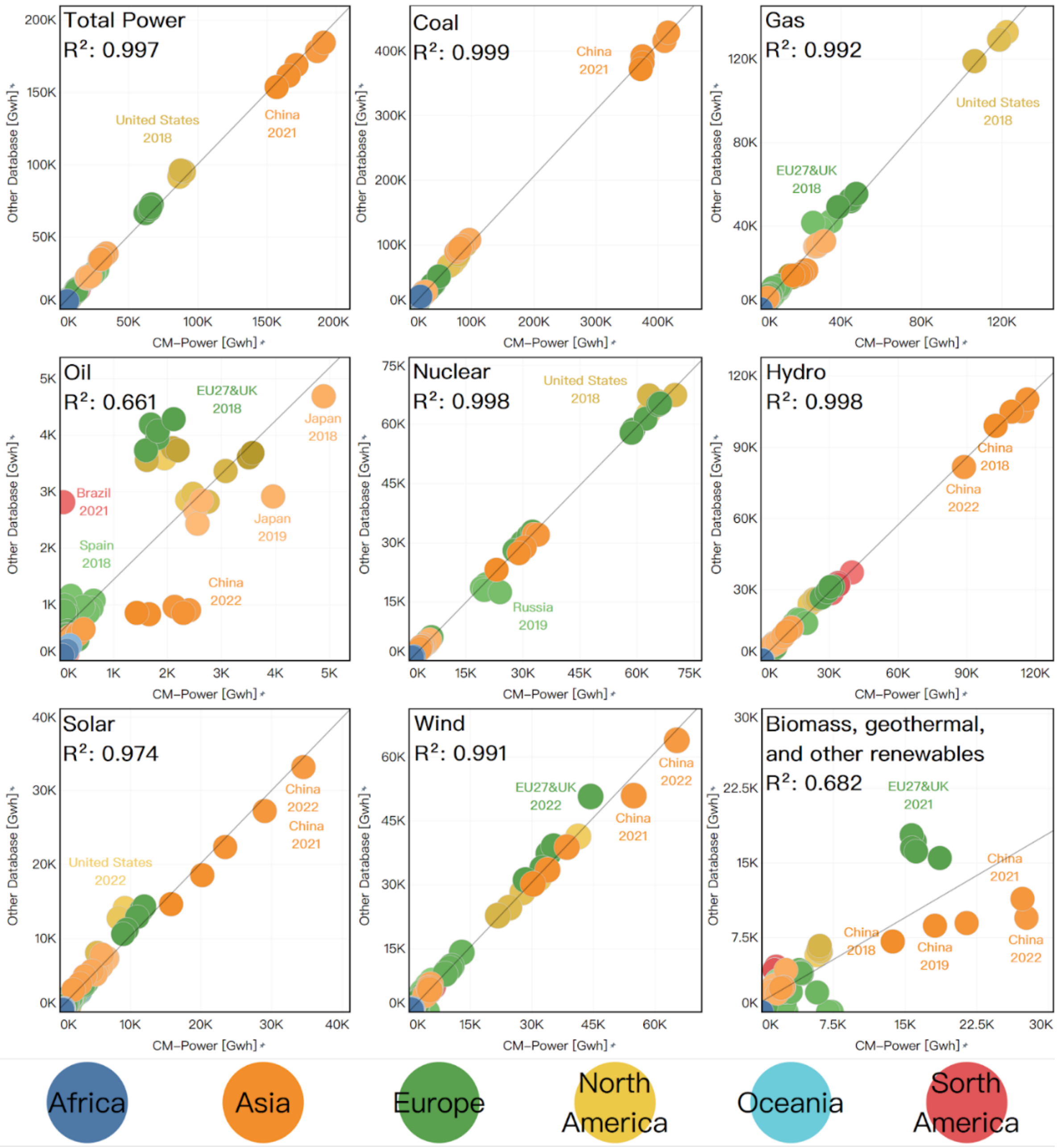

**Fig 3 | Correlation between Carbon Monitor-Power power generation data and power generation from other databases (BP[2] for Russia and South Africa and IEA[20] for the rest countries).** Different colors indicate different world regions. The correlation coefficient is shown as squared term $R^2$.

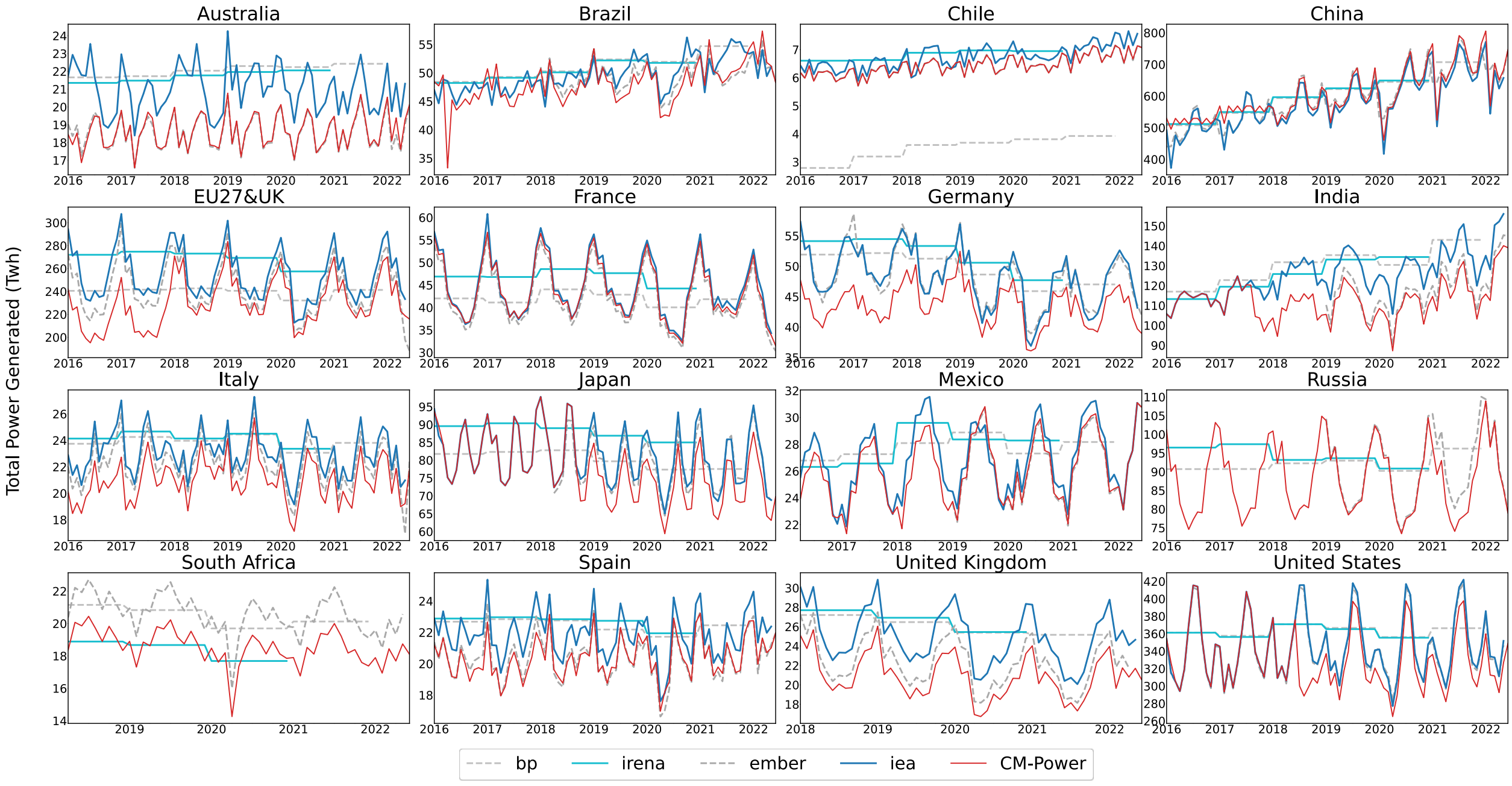

**Fig 4 | Time series of monthly total power generation data from Carbon Monitor-Power (red lines) and reference data.** Dashed light gray lines for BP's Statistical Review of World Energy[2], solid light blue lines for The International Renewable Energy Agency (IRENA)'s Renewable Energy Statistics[19], dashed dark grey lines for Ember's monthly power generation[31], and solid dark blue lines for IEA's monthly electricity statistics[20]). For data from BP and IRENA, as the original data are provided as an annual total, it is plotted as a monthly average for comparison purposes.

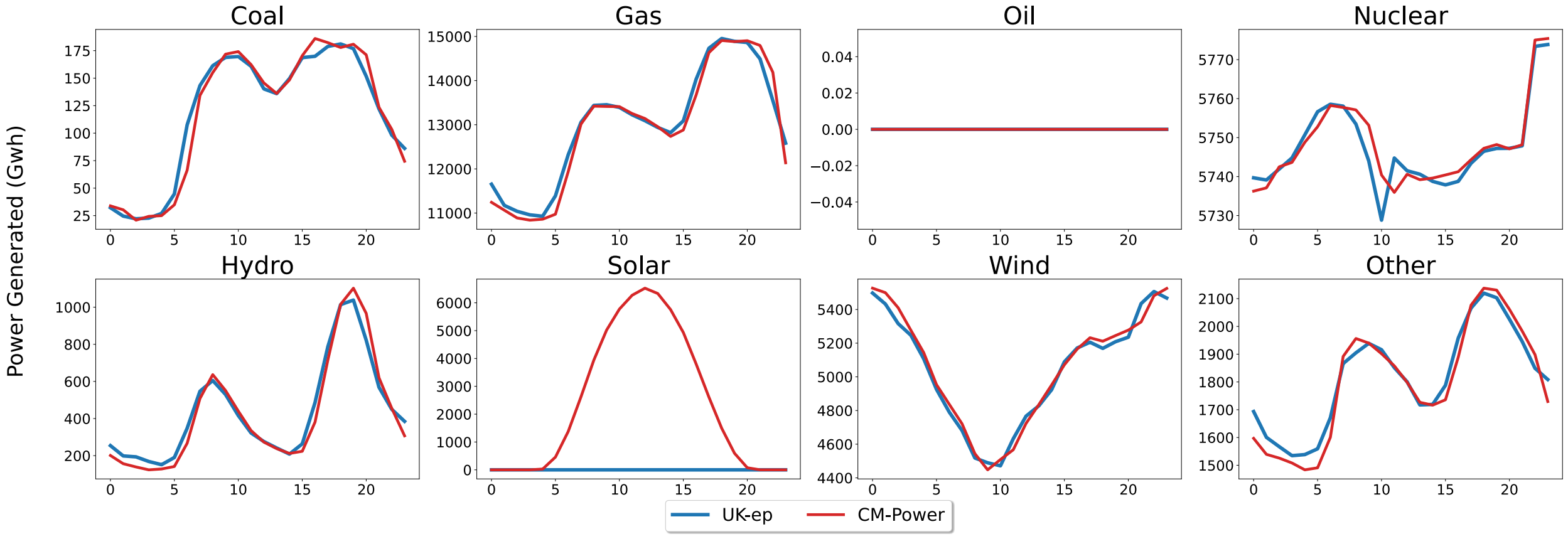

**Fig 5 | Comparison between two databases: Carbon Monitor-Power (red) and UK_ep (blue) for eight energy sources: Coal, Gas, Oil, Nuclear, Hydro, Solar, Wind, and Other Renewables.** The x-axis denotes the hour of the day and the Y-axis presents and power generated from each source. The figure shows an average hourly profile for the month of June 2022.

# Tables

**Table 1 | Summary table of power generation data sources, generation types included in original data, and their spatial and temporal resolutions.**

| Country/Region | Data source | Energy sources included | Temporal Resolution | Spatial Resolution |
|---|---|---|---|---|
| Australia | OpenNEM (https://opennem.org.au/energy/nem/?range=7d&interval=30m) | Solar (Rooftop), Solar(utility), Wind, Hydro, Gas (Waste Coal Mine), Gas (Reciprocating), Gas (OCGT), Gas (CCGT), Gas (Steam), Distillate (Energy source: Diesel), Bioenergy (Biomass), Bioenergy (Biogas), Coal (Black), Coal (Brown) | Hourly | National and sub-national* |
| Brazil | Operator of the National Electricity System (http://www.ons.org.br/Paginas/) | Coal, Natural-gas, Petroleum, Wind, Solar, Hydro, Other | Hourly | National and sub-national* |
| China | National Bureau of Statistics (https://data.stats.gov.cn/); China Electricity Council (https://cec.org.cn/); CCTD (https://www.cctd.com.cn) | Thermal, Coal, Natural-gas, Hydro, Solar, Wind, Nuclear | Daily (CCTD) / Monthly (NBS and CEC) | National |
| EU27 (including all EU countries except for Malta) | ENTSO-E (https://www.entsoe.eu) | Coal (Brown coal, coal derived gas, hard coal, peat), Natural-gas, Oil (oil and shale oil), Hydro (pumped storage, run-of-river, water reservoir), Solar, Wind, Nuclear, biomass, geothermal, other renewables, waste, and other) | Hourly/Sub-hourly | National |
| UK | BMRS (https://www.bmreports.com/bmrs/) Sheffield Solar (https://www.solar.sheffield.ac.uk/) | Coal, Natural gas (combined cycle gas turbine, ccgt), Natural gas (open gas cycle turbine, ocgt), Oil, Nuclear, Hydro, Wind, Biomass, Other, and Solar | Sub-hourly | National |
| India | Power System Operation Corporation Limited (https://posoco.in/reports/daily-reports/) | Coal, Lignite, Hydro, Nuclear, Gas_Naptha_Diesel, RES (including Wind, Solar, Biomass and Others) | Daily | National / sub-national* |
| Japan | Organization for Cross-regional Coordination of Transmission Operators (OCCTO) (https://www.occto.or.jp/en/) | Nuclear, Geothermal, Hydroelectric, Fossil fuels, Photovoltaic, Pumped Storage Hydroelectricity, Wind, Wind regulated | Hourly | National/ sub-national** |
| Russia | United Power System of Russia (http://www.so-ups. ru/index.php) | Thermal, Nuclear, Solar, Hydro, Renewables | Hourly | National |
| South Africa | Eskom (https://www.eskom.co.za/dataportal/supply-side/station-build-up-for-the-last-7-days/) | Coal (labeled as Thermal in the source data), Natural-gas, Oil (labeled as OCGT in the source data), Nuclear, Pumped Water Generation, Hydro Water Generation, Photovoltaic generation (PV), Concentrated Solar Power generation (CSP), Wind, Other | Hourly | National |
| United States | Energy Information Administration's (EIA) Hourly Electric Grid Monitor (https://www.eia.gov/beta/electricity/gridmonitor/) | Coal, Natural-gas, Petroleum, Nuclear, Hydro, Solar, Wind, Other | Hourly | National/ sub-national* |
| Mexico | Gobierno De Mexico (https://www.cenace.gob.mx/Paginas/SIM/Reportes/EnergiaGeneradaTipoTec.aspx) | Coal, Gas, Combined cycle, Internal Combustion, Conventional Thermal, Nuclear, Hydro, Wind, Solar, Biomass, Geothermal | Hourly | National |
| Chile | Coordinador Eléctrico Nacional (https://www.coordinador.cl/operacion/graficos/operacion-real/generacion-real/) | Coal, Petcoke, Biogas, Natural Gas, Diesel, Fuel Oil, Run-of-river, Storage, Wind, Solar, Geothermal, Biomass, Cogeneration | Hourly | National |

* This 'sub-national' represents the administrative unit

** This 'sub-national' represents utility company

**Table 2. Summary of power generation datasets characteristics and comparison statistics including coefficient of determination ($R^2$) mean relative difference (Rd), and sample size (n) when compared with Carbon Monitor-Power**

| **Dataset** | Carbon Monitor-Power | IEA | BP | IRENA | Ember |
|---|---|---|---|---|---|
| **Spatial coverage** | 37 countries including | 47 countries | Global | Global | 85 geographies |
| **Temporal coverage** | 2016-2022 | 2000-2022 | 1985-2021 | 2012-2021 | 2018-2022 |
| **Temporal resolution** | Daily and hourly | Monthly | Annual | Annual | Monthly |
| **Latency** | 1 day (except for China which has 3 weeks latency) | 2~3 months | 8 months | 7 months | 1 month |
| **Method** | Generation data acquired from national electricity grids, aggregated/ disaggregated to hourly/daily resolution§§ | Monthly reporting by country | Primary official sources and third-party data[2] | Data acquired through a combination of methods: IRENA questionnaire, official national statistics, industry association reports, consultant reports and news articles | National grid/national statistics data when available, combined with other data sources including IEA, IRENA |
| **Comparison with Carbon Monitor-Power** | | | | | |
| **Total generation** | / | $R^2$ = 0.99<br>Rd = 2%<br>n = 38 | $R^2$ = 0.99<br>Rd = 5%<br>n = 40 | $R^2$ = 0.99<br>Rd = 7%<br>n = 38 | $R^2$ = 0.99<br>Rd = 1%<br>n = 44 |
| **Coal** | / | $R^2$ = 0.99<br>Rd = 1%<br>n = 29 | $R^2$ = 0.98<br>Rd = 6%<br>n = 16 | / | $R^2$ = 0.99<br>Rd = 2%<br>n = 33 |
| **Gas** | / | $R^2$ = 0.98<br>Rd = 10%<br>n = 35 | $R^2$ = 0.95<br>Rd = 23%<br>n = 15 | / | $R^2$ = 0.99<br>Rd = 11%<br>n = 38 |
| **Oil** | / | $R^2$ = 0.79<br>Rd = 22%<br>n = 22 | $R^2$ = 0.71<br>Rd = 14%<br>n = 14 | / | $R^2$ = 0.79<br>Rd = 58%<br>n = 24 |
| **Nuclear** | / | $R^2$ = 0.99<br>Rd = -1%<br>n = 21 | $R^2$ = 0.98<br>Rd = 1%<br>n = 24 | / | $R^2$ = 0.99<br>Rd = -1%<br>n = 24 |
| **Hydro** | / | $R^2$ = 0.97<br>Rd = 2%<br>n = 34 | $R^2$ = 0.91<br>Rd = --1%<br>n = 36 | $R^2$ = 0.92<br>Rd = 0%<br>n = 34 | $R^2$ = 0.97<br>Rd = -5%<br>n = 39 |
| **Wind** | / | $R^2$ = 0.99<br>Rd = 2%<br>n = 36 | $R^2$ = 0.96<br>Rd = 1%<br>n = 38 | $R^2$ = 0.95<br>Rd = 0%<br>n = 36 | $R^2$ = 0.99<br>Rd = 1%<br>n = 41 |
| **Solar** | / | $R^2$ = 0.97<br>Rd = 7%<br>n = 31 | $R^2$ = 0.93<br>Rd = 10%<br>n = 34 | $R^2$ = 0.93<br>Rd = 7%<br>n = 32 | $R^2$ = 0.96<br>Rd = 9%<br>n = 33 |
| **Biomass, Geothermal and other Renewables** | / | $R^2$ = 0.68<br>Rd = -14%<br>n = 34 | $R^2$ = 0.65<br>Rd = -16%<br>n = 36 | $R^2$ = 0.50<br>Rd = -24%<br>n = 34 | $R^2$ = 0.75<br>Rd = -32%<br>n = 36 |